\def\year{2017}\relax
\documentclass[letterpaper] {article} %DO NOT CHANGE THIS
\pdfoutput=1
\usepackage{hyperref}
\hypersetup{
  pdfinfo={
    Title={Prototype Tasks: Improving Crowdsourcing Results through Rapid, Iterative Task Design},
    Author={Neil Gaikwad},
    Subject={HCOMP AAAI},
  }
}
\usepackage{pdfpages}
\usepackage{aaai17}  %Required
\usepackage{times}  %Required
\usepackage{helvet}  %Required
\usepackage{courier}  %Required
\usepackage{url}  %Required
\usepackage{graphicx}  %Required
\usepackage{siunitx}
\usepackage{comment}
\usepackage{float}
\usepackage[skip=8pt]{caption}

\begin{document}

% The file aaai.sty is the style file for AAAI Press 
% proceedings, working notes, and technical reports.
%
\title{Prototype Tasks: Improving Crowdsourcing Results through \\ Rapid, Iterative Task Design}
\author{\\
Snehalkumar `Neil' S. Gaikwad, Nalin Chhibber, Vibhor Sehgal, Alipta Ballav, Catherine Mullings, Ahmed Nasser, \\ 
Angela Richmond-Fuller, Aaron Gilbee, Dilrukshi Gamage, Mark Whiting, Sharon Zhou, Sekandar Matin, Senadhipathige Niranga, \\
Shirish Goyal, Dinesh Majeti, Preethi Srinivas, Adam Ginzberg, Kamila Mananova, Karolina Ziulkoski, Jeff Regino, Tejas Sarma, \\ Akshansh Sinha, Abhratanu Paul, Christopher Diemert, Mahesh Murag, William Dai, Manoj Pandey, Rajan Vaish and Michael Bernstein \\ \\
Stanford Crowd Research Collective \\ 
daemo@stanford.edu \\ 
\\
}
 
\maketitle
\begin{abstract}
Low-quality results have been a long-standing problem on microtask crowdsourcing platforms, driving away requesters and justifying low wages for workers. To date, workers have been blamed for low-quality results: they are said to make as little effort as possible, do not pay attention to detail, and lack expertise. In this paper, we hypothesize that \textit{requesters} may also be responsible for low-quality work: they launch unclear task designs that confuse even earnest workers, under-specify edge cases, and neglect to include examples. We introduce \textit{prototype tasks}, a crowdsourcing strategy requiring all new task designs to launch a small number of sample tasks. Workers attempt these tasks and leave feedback, enabling the requester to iterate on the design before publishing it. We report a field experiment in which tasks that underwent prototype task iteration produced higher-quality work results than the original task designs. With this research, we suggest that a simple and rapid iteration cycle can improve crowd work, and we provide empirical evidence that requester ``quality'' directly impacts result quality.
\end{abstract}

\subsection{\centering Introduction}

Quality control has been a long-standing concern for microtask crowdsourcing platforms such as Amazon Mechanical Turk (AMT)~\cite{ipeirotis2010quality,marcus2015crowdsourced}. Low quality work causes requesters to mistrust the data they acquire~\cite{kittur2008crowdsourcing}, offer low payments~\cite{ipeirotis2010mechanical}, and even leave the platform in frustration. Currently, the dominant narrative is that low-quality work is the fault of \textit{workers}---those who demonstrate ``a lack of expertise, dedication [or] interest''~\cite{sheng2008get}, or are either ``lazy'' or ``overeager''~\cite{bernstein2010soylant}. Workers are blamed for cheating~\cite{christoforaki2014step,martin2014being}, putting in minimal effort~\cite{bernstein2010soylant,rzeszotarski2011instrumenting}, and strategically maximizing earnings~\cite{kittur2008crowdsourcing}. To counter such behavior, requesters introduce worker-filtering strategies such as peer review~\cite{dow2012shepherding}, attention checks with known answers~\cite{le2010ensuring}, and majority agreement~\cite{von2004labeling}. Despite these interventions, the quality of crowd work often remains low~\cite{kittur2013future}. 

This narrative ignores a second party who is responsible for microtask work: the \textit{requesters} who design the tasks. Since HCOMP's inception in 2009, many papers have focused on quality control by measuring and managing crowd workers --- however, very few have implicated requesters and their task designs in quality control. For example, a system might give workers feedback on their submission, while taking the requesters’ tasks as perfect and fixed~\cite{dow2012shepherding}. This assumption is even baked into worker tools such as Turkopticon~\cite{irani2013turkopticon}, which include ratings for fairness and payment but not task clarity. \textit{Should requesters also bear responsibility for low-quality results?}

Human-computer interaction has long held that errors are the fault of the designer, not the user. Through this lens, the designer (requester) is responsible for conveying the correct mental model to the user (worker). There is evidence that requesters can fail at this goal, forgetting to specify edge cases clearly~\cite{martin2014being} and writing text that is not easily understood by workers~\cite{khanna2010evaluating}. Requesters, as domain experts, underestimate task difficulty~\cite{hinds1999curse} and make a fundamental attribution error~\cite{ross1977intuitive} of blaming the person for low-quality submissions rather than the task design that led them to produce those results. Recognizing this, workers become risk-averse and avoid requesters who could damage their reputation and income~\cite{martin2014being,mcinnis2016taking}. 

Workers and requesters share neither the same expectations for results nor their objectives for tasks~\cite{kittur2008crowdsourcing,martin2014being}. For instance, requesters may feel that the task is clear, but that workers do not put in sufficient effort~\cite{bernstein2010soylant,kittur2008crowdsourcing}, thus rejecting work that they judge to be of low quality. Workers, on the other hand, may feel that requesters do not write clear tasks~\cite{mcinnis2016taking} or offer fair compensation~\cite{martin2014being}. Overall, requesters' mistakes lead workers to avoid tasks that appear too confusing or ambiguous in order to avoid getting rejected and harming their reputation~\cite{irani2013turkopticon,mcinnis2016taking}. 

Rather than improving result quality by designing crowdsourcing processes for workers~\cite{mcinnis2016taking,dow2012shepherding}, in this paper we capitalize on the notion of broken windows theory~\cite{wilson1982broken} and improve quality by designing crowdsourcing processes for requesters. Our research contributes: 1) an integration of traditional human-computer interaction techniques into crowdsourcing task design via prototype tasks, and, through an evaluation, 2) empirical evidence that requester ``quality'' directly impacts task quality.

\subsection{\centering Prototype Task System}

While prior work has studied the impact of enforcement mechanisms such as attention checks~\cite{kittur2008crowdsourcing,mitra2015comparing}, our focus is on requesters' ability to build common ground~\cite{clark1991grounding} with workers through instructions, examples, and task design. Inspired by the iterative design process in human-computer interaction~\cite{preece1994human,sharp2007interaction,harper2008human,holtzblatt2004rapid}, we incorporate worker feedback in the task design process. We build on the prototyping and evaluation phases of user-centered design~\cite{baecker2014readings,dix2009human,holtzblatt2004rapid,nielsen2000you} and introduce \textit{prototype tasks} as a strategy for crowdsourcing task design \textit{(Appx. B)}. Prototype tasks, in-progress task design, draw on an effective practice in crowdsourcing of identifying high-quality workers by posting small tasks to vet them first~\cite{mitra2015comparing,horeFast}.  

Prototype tasks, integrated into the Daemo crowdsourcing marketplace~\cite{gaikwad2015daemo}, require that all new tasks go through a feedback iteration from 3--7 workers on a small set of inputs before being launched in the marketplace. These workers attempt the task and leave feedback on the task design. First, prototype tasks involve adding a feedback form to the bottom of the task. This might be a single textbox, or following recent design thinking feedback practice~\cite{dschool}, a trio of ``I like...'', ``I wish...'' and ``What if...?'' feedback boxes. Second, to launch a prototype task, a requester samples a small number of inputs (e.g., a few restaurants to label), and launches them to a small number of workers. Third, they utilize the results and feedback to iterate on their task design \textit{(Appx. B)}. 

We argue that prototype tasks, which take only a few minutes and provide a ``thin slice'' of feedback from a few workers, are enough to identify edge cases, clarify instructions, and engage in perspective-taking as a worker. \textit{We hypothesize that this iteration will result in better designed tasks, and downstream, better results.}

\subsection{\centering Evaluation and Results}
Prototype tasks embody a proposition that requesters’ task design skills directly impact the work quality they receive, and that processes from user-centered design might produce higher-quality task designs. To investigate the claim, we recruited  7 experienced microtask crowdsourcing requesters. Requesters each designed three different canonical types of prototype tasks detailed in prior work~\cite{Justin2015ETA}, then launched them to a small sample of workers on AMT for feedback, and reviewed the results to iterate and produce revised task designs. The types of the tasks were chosen based on the common crowdsourcing microtask categories: categorization, search, and description \textit{(Appx. A)}. Each task was known to be either susceptible to misinterpretation or a challenging edge case. We provided requesters with a short description of the tasks they should implement and private ground-truth results that they should replicate. We then launched both the original and the revised tasks on AMT, randomizing workers between the two task designs. Requesters reviewed pairs of results blind to condition and chose the set they found to be higher quality. If workers are primarily responsible for low-quality crowd work, then the original and revised tasks would both have equally poor results. However, if the poor quality is due at least in part to the requesters, then the prototype tasks would lead to higher-quality results.  

We found that the prototype task intervention produced results that were preferred by requesters significantly more often than the original design. Requesters preferred the results from the Prototype (revised) tasks nearly two thirds of the time (51 vs. 31). This result was significant ($\chi^2(1)$=4.878, $p<0.05$). In other words, modifications to the original task design and instructions had a positive impact on the final outcome. Thus, our hypothesis was supported: result quality is at least in part attributable to requesters' task designs. The requesters themselves did not change, and the distribution of worker quality was identical due to randomization, but improved task designs led to higher-quality results. While only seven participants limits generalizability, the significant result with just seven is an indirect statement of the large effect size of the intervention. Future work involving the longitudinal experiments of tasks will help better triangulate the phenomenon and dynamics of the requesters' and workers' impact on quality.
 
%We also found that all the tasks had similar patterns and no particular task dominated the others in the analysis. 

To understand where requesters might misstep, we next investigated the written feedback that workers gave during the first phase, and the changes that requesters made to their tasks in response to the feedback. We conducted a grounded analysis using workers' feedback on flaws in tasks' designs and instructions. Two researchers independently coded each item of feedback and each task's design changes according to inductively derived categories from data, resolving disagreements through discussion. Raw Cohen's $\kappa$ agreement scores for these categories ranged from 0.64--1.0, indicating strong agreement. This process generated four categories of feedback and iteration: unclear/vague instructions, missing examples, incorrect grammar, and broken task UI/accessibility \textit{(Appx. B)}. The most common worker feedback was on the user interface and accessibility of the task (49\% of feedback). %These fall into the categories of feedback that would often arise through heuristic evaluation and usability studies: for example, making links clickable, confusing interface elements, and broken or inaccessible links. %Requesters were open to redesigning low-level interface elements: 60\% of revised tasks included user interface amendments.

%Our experiment involved real-world tasks, real crowd workers and requesters from crowdsourcing platforms. Prototype tasks generate a large amount of training data for a machine learning system to learn over time how tasks evolve to become more effective. With some task design challenges, it may be possible to flag tasks that do not follow best practices and warn the requester. We plan to provide an automated analysis of task designs.

\subsection{\centering Conclusion}
%Effective tasks often include several important features as indicated by workers' feedback: a summary description, example inputs and outputs, and a description of what to do with edge cases.  
In this paper, we introduce prototype tasks, a strategy for improving crowdsourcing task results by gathering feedback on a small sample of the work before launching. We offer empirical evidence that requesters' task design directly impacts result quality. Based on this result, we encourage requesters to take ownership of the negative effects they may have on their own results. This work ensures that the discussion of crowdsourcing quality continues to include requesters' behaviors and emphasize how their upstream decisions can have downstream impacts on work quality. 

\newpage
\clearpage

\bibliographystyle{aaai}
\bibliography{Prototype}

\onecolumn

\subsection{\centering Appendix A: Tasks}

\vspace{2em}
\textbf{Three tasks the requresters designed}  
 
\begin{itemize}
 \item \textit{Marijuana legalization (classification)}~\cite{metaxa2016web}: Requesters designed a task to determine ten websites' stances on marijuana legalization (Pro-legalization, Anti-legalization, or Neither). We chose this task because it captures a common categorical judgment task with forced-choice selection: other examples include image labeling~\cite{deng2009imagenet} and sentiment classification~\cite{callison2010creating}. Similar to many such tasks, this one includes several edge cases, for example whether an online newspaper that neutrally reports on a pro-legalization court decision is expressing a `Pro-legalization' or `Neither' stance. Marijuana Legalization task included 10 URLs.

\item \textit{Query answering (search)}~\cite{bernstein2012direct}: Requesters were given a URL with a set of search queries that led to the URL and designed a task asking workers to copy-paste text from each webpage (URL) that answers those queries. One such query was ``dog body temperature'' (correct response: 102.5\si{\degree}$F$). We chose this task to be representative of search and data extraction tasks on Mechanical Turk~\cite{Justin2015ETA}. The challenge in designing this task correctly is in providing careful instructions about how much information to extract. For instance, without careful instruction, workers would often conservatively capture a large block of text instead of just the information needed~\cite{bernstein2012direct}. Query answering (search) included 8 search task URLs.

\item \textit{Video summarization (description)}~\cite{kim2016storia,wu2011video,lasecki2014legion}: Requesters were given a YouTube video and requested to design a task for workers to write a text summary of the video. This was the most open-ended of the three tasks, requiring free-text content generation by the workers. To achieve a high-quality result, the task design needed to specify desired length, style, and example high and low quality summaries of a complex concept Video Summarization included 1 URL.  
\end{itemize}

\vspace{5em}
\subsection{\centering Appendix B: Figures}
\begin{figure*}[th]
\centering
\caption{Workers provided feedback on tasks that ranged from clarity to payment and user interface}

  \includegraphics[width=1.0\columnwidth]{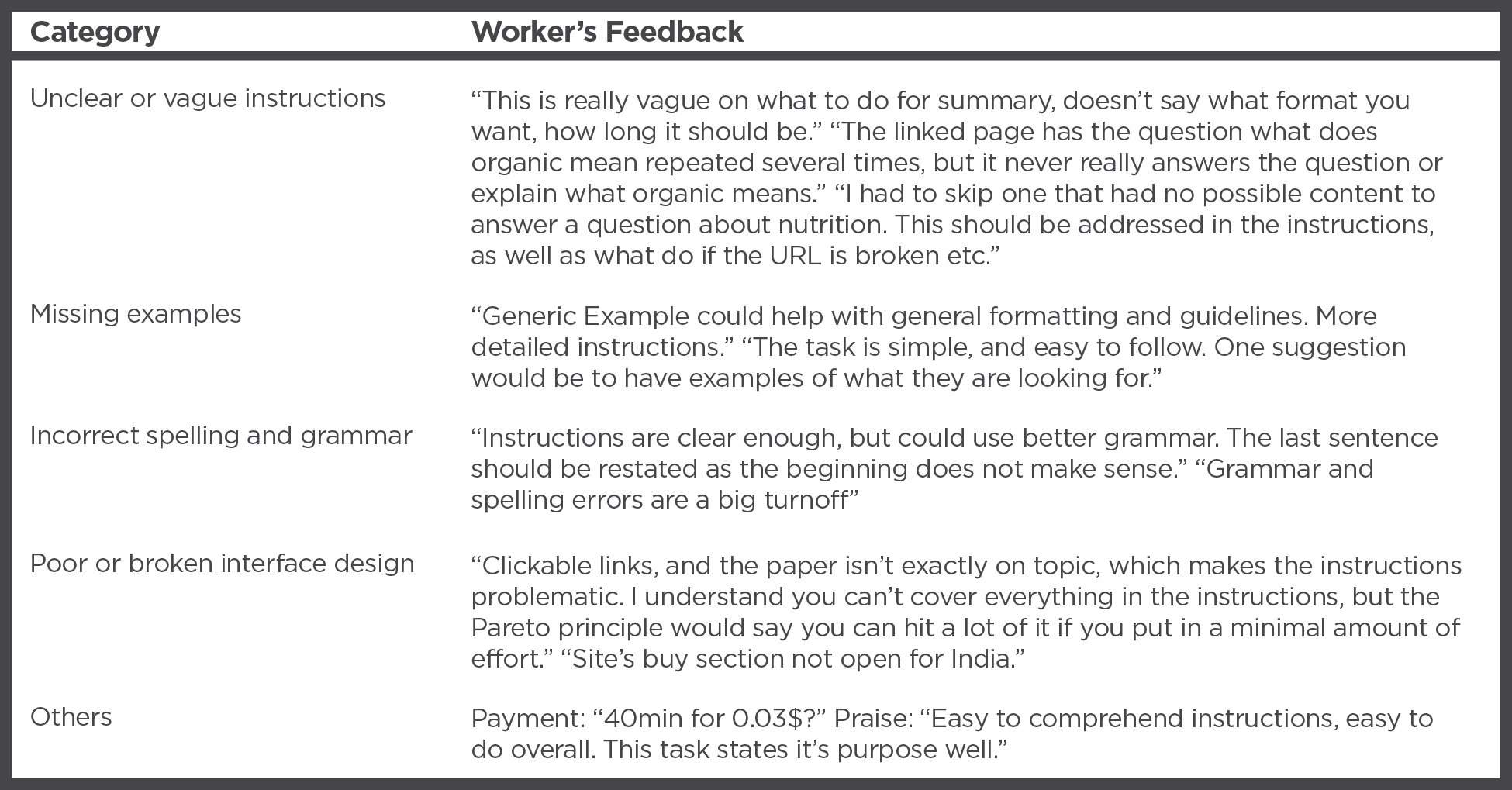}
\label{fig:workerFeedback}
\end{figure*}

\twocolumn

\begin{figure*}[ht]
\centering
\caption{With \textit{prototype tasks}, all tasks are first launched to a small sample of workers to gather free-text feedback and example results. Requesters use prototype tasks to iterate on their tasks before launching to the larger marketplace.}
\includegraphics[width=0.8\columnwidth]{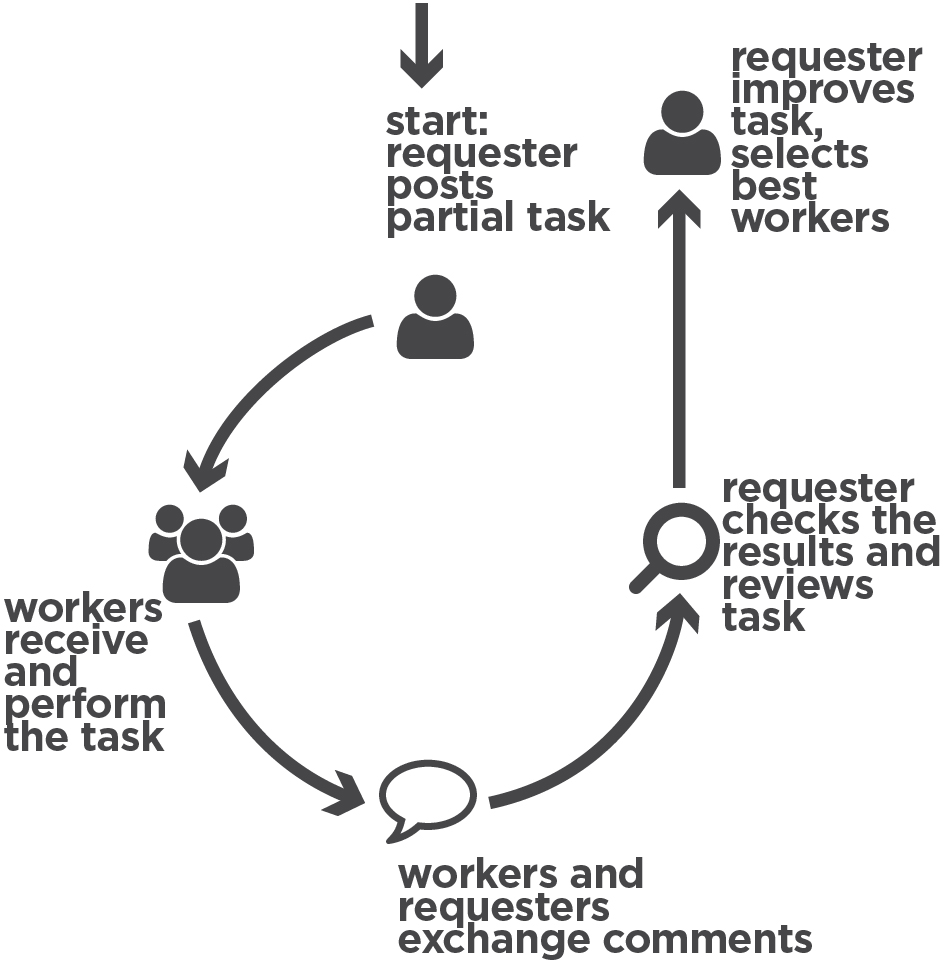}
\label{fig:workflow}
\end{figure*}

\begin{figure*}[ht]
\centering
\caption{Daemo's prototype task feedback from workers (right) can be used to revise the task interface (left). Requesters can review both the written feedback and the submitted sample results.}~\label{fig:taskFeedback}

\includegraphics[width=1.7\columnwidth]{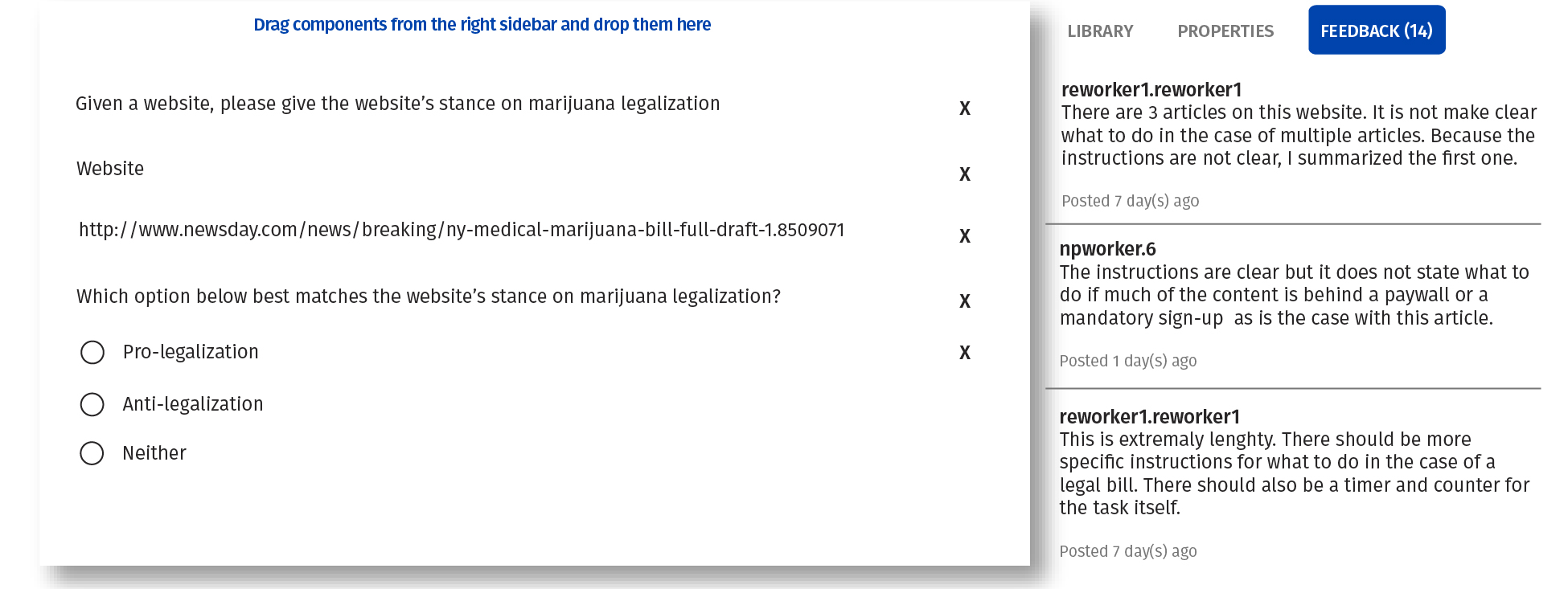}
\end{figure*}

\begin{figure*}[t]
\centering
\caption{Prototype tasks appear alongside other tasks in workers' task feeds on the Daemo crowdsourcing platform. Prototype tasks are paid tasks in which 3--7 workers attempt a task that will soon go live on the platform and give feedback to the requester to inform an iterative design cycle.}
  \includegraphics[width=1.5\columnwidth]{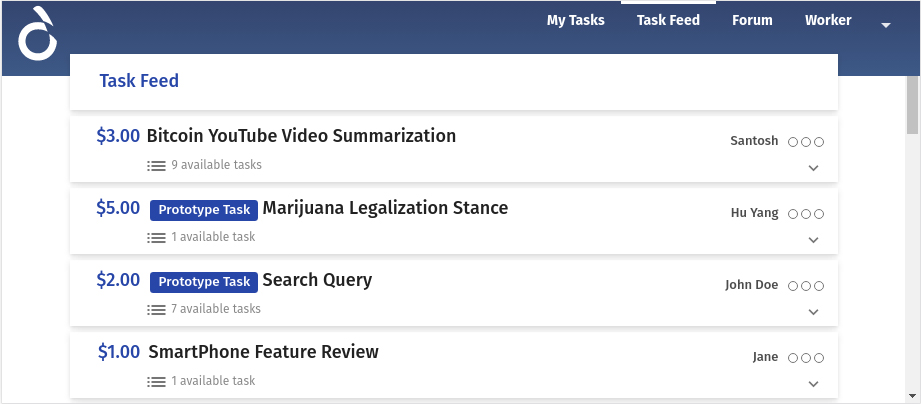}
\label{fig:taskFeed}
\end{figure*}

\end{document}